# Transformation of the mechanical properties of materials by the geomagnetic resonance

Vladimir I. Alshits[1], Elena V. Darinskaya[1], Marina V. Koldaeva[1], Romuald K. Kotowski*[2] and Elena A. Petrzhik[1]

[1]Department of Crystallography and Photonics of FSRC, Shubnikov Institute of Crystallography of RAS, Leninskii prosp. 59, Moscow 119333, Russia
[2]Faculty of Information Technology, Polish-Japanese Academy of Information Technology, Koszykowa 86, 02-008 Warsaw, Poland

E-mail: rkotow@pja.edu.pl

**Abstract.** The strong geomagnetic influences on mechanical properties of crystals due to their exposure to ultralow crossed magnetic fields, the Earth's field (~50 µT) and the AC field (~3 µT) in the electron paramagnetic resonance scheme, are discussed. Resonance relaxation displacements of dislocations in *NaCl* crystals are found both for a harmonic pump field and for a pulse AC field of resonance duration ~0.5 µs. Resonant changes have been also detected in the microhardness of *ZnO*, triglycine sulfate, and potassium acid phthalate crystals after their exposure in the same EPR scheme. The both effects manifest new strongly anisotropic properties. In particular, the frequency of the resonance is very sensitive to the mutual orientation of the sample and the Earth's field. Physical mechanisms and practical significance of the phenomenon are discussed.

**Keywords:** magnetoplasticity, resonance, properties of materials, EPR, magnetic fields

## 1. INTRODUCTION

The magnetoplastic effect (MPE) was discovered as the movement of dislocations in *NaCl* crystals held in a DC magnetic field $B \approx 0.5$ T in the absence of a mechanical load Alshits (1987). Numerous subsequent independent studies (see for example the review papers by Alshits (2003); Golovin (2004), Morgunov (2004), Alshits (2008)), showed that this phenomenon is due to a magnetically induced change in the structure of impurity stoppers on dislocations, which results in the reduction of the force of pinning of dislocations and their relaxation displacements in the field of the internal stresses. The role of the magnetic field is reduced to changing the spin states of the impurity centers with the removal of the quantum exclusion of a certain electronic transition, which leads to their transformation.

The phenomenon manifests itself also in macroplasticity of crystals: under the active loading when the stress load is increasing with the constant rate: $\dot{\sigma} = $ const Golovin (1995), active deformation $\dot{\varepsilon} = const$ Urusovskaya (2003), creep $\sigma = $ const Smirnov (2001), and internal friction Tyapunina (1999).

The magnetoplastic effect is observed in crystals of various types not only in a static magnetic field but also in crossed static and alternating magnetic fields in the electron paramagnetic resonance (EPR) regime at the resonance frequency

$$\nu_r = g\mu_B B/h, \qquad (1)$$

where the $g$-factor is usually close to 2, $\mu_B$ is the Bohr magneton, and $h$ is the Planck's constant. Such resonances in *NaCl* and silicon crystals were found by the three independent groups at the frequency of 9.5 GHz and the related magnetic field ~0.3 T Golovin (1998), (2000), (2002), Osip'yan (2004) and Badylevich (2005).

More recently, a similar resonance in *NaCl* was detected and studied in ultralow fields: the Earth's field ~50 $\mu$T and the pump field with an amplitude of 3 $\mu$T and a frequency of ~1 MHz Alshits (2016). This particular low-frequency resonance will be a subject of this paper. As we shall see, it has a number of new very informative properties, primarily the strong anisotropy of the effect with respect to the mutual orientation of the crystal, dislocations, and magnetic fields. In particular, the resonance frequency becomes very sensitive to the angle $\theta$ made by the sample with the Earth's magnetic field (giant anisotropy of the effective $g$-factor: $g_{eff} \approx g_0 \cos\theta$). In addition, as will be shown, this resonance also manifests itself in changing the microhardness of the zinc oxide (*ZnO*), triglycine sulfate (*TGS*), and potassium acid phthalate (*KAP*) crystals after their exposure in the same EPR scheme in the Earth's magnetic field together with very strong anisotropy of the effect including even the giant anisotropy of $g$-factor.

## 2. EXPERIMENT

To create the EPR conditions, the samples were placed in crossed static and alternating magnetic fields. In most experiments as the static magnetic field the Earth's magnetic field was used. Its direction and magnitude were measured at the place of a sample. The vector of the Earth's magnetic field $\mathbf{B}_{Earth}$ made an angle of 29.5° with the vertical direction and had a length of 49.97 $\mu$T. The alternating pump field $\tilde{\mathbf{B}}$ was produced in a screened coaxial chamber around rectilinear conductor *1* through which a sinusoidal current flowed as shown in Fig. 1a. Its frequency $\nu$ varied in the range of 10 kHz to 1.5 MHz. The amplitude of the harmonic pump field was from 1 to 6 $\mu$T and the time of exposure was from 15 s to 2 h depending on the type of the tests.

In situ effects were studied when the response of a crystal to the magnetic action occurs in the process of exposure and directly involves moving dislocations. "Memory" effects were also studied when the response occurs in a certain time after exposure. They were found by measuring the microhardness of *ZnO*, *TGS*, and *KAP* crystals.

In the former case, histograms of paths of fresh dislocations in *NaCl* crystals introduced in the samples just before the exposure were measured. Using these histograms, the mean free paths $l$ for various physical parameters and conditions was calculated. Their positions before and after exposure were determined by the selective etching method. In addition to the absolute mean path $l$, the normalized dimensionless path $l\sqrt{\rho}$ which is the ratio of the $l$ to the average distance $1/\sqrt{\rho}$ between dislocations was used. Here $\rho \sim 10^4$ cm$^{-2}$ is the total density of dislocations.

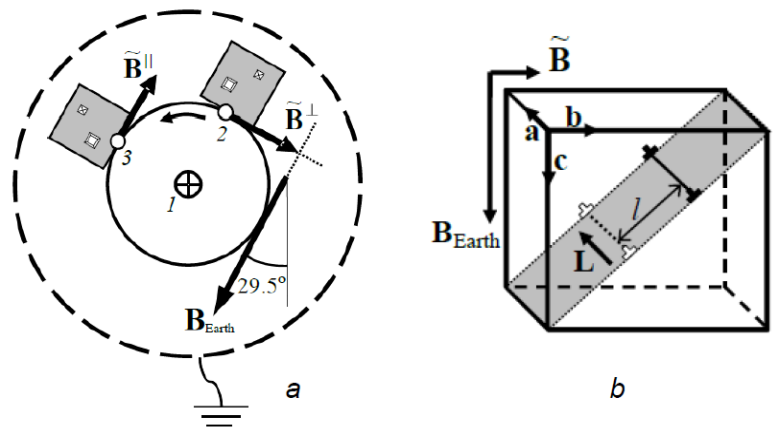

Fig. 1 (a) Scheme of the experiment in the crossed magnetic fields, (b) the geometry configuration of dislocation motion in the *NaCl* crystal.

## 3. RESONANCE OF DISLOCATION DISPLACEMENTS

*3.1 Basic features of the harmonic resonance*

Effect of the mutual orientation of dislocations and magnetic fields

The *NaCl* samples were chipped out along the {100} cleavage planes and had approximate dimensions of 3×3×5 mm. Fresh dislocations were introduced in a sample by a weak impact. In this case, most dislocations were rectilinear being directed along the directions **L** ∥ **a** = [100], **b** = [010], or **c** = [001]. Their slip planes belonged to the {110} system. Here we described the experiments with the sample oriented so that the edge **c** was parallel to the Earth's field and the **a** edge was orthogonal to the plane of the magnetic fields {$\mathbf{B}_{Earth}$, $\widetilde{\mathbf{B}}$} (see Fig. 1b). In this geometry of crossed magnetic fields, **a**-dislocations (**L** ∥ **a**) are the most mobile whereas **b**- and **c**-dislocations have noticeably smaller paths under the same *conditions*. The orientation **L** ∥ **a** ⊥ {$\mathbf{B}_{Earth}$, $\widetilde{\mathbf{B}}$} the chosen position of the sample with respect to the magnetic fields remains optimal for all studied crystals *NaCl*.

The intensity of the discussed resonance depends also on the mutual orientation of the fields $\mathbf{B}_{Earth}$ and $\widetilde{\mathbf{B}}$. However, this type of anisotropy is not universal and is manifested in different degrees in crystals with different impurity compositions. Figure 1a shows two positions *2* and *3* of the sample that correspond to the mutually orthogonal and parallel fields: $\widetilde{\mathbf{B}} = \widetilde{\mathbf{B}}^{\perp} \perp \mathbf{B}_{Earth}$ and $\widetilde{\mathbf{B}} = \widetilde{\mathbf{B}}^{\parallel} \parallel \mathbf{B}_{Earth}$. Normally, for usual EPR with high-frequency pump field (~10 GHz) the first of these orientations is much more preferable. In our case of low-frequency EPR (~1 MHz) we met in *NaCl* crystals with different impurities both traditional situation and the cases of comparable dislocation responses. Figure 2a demonstrates the example of a traditional case which is realized in *NaCl* crystals with *Ni* impurities where the pump field $\widetilde{\mathbf{B}}^{\parallel}$ causes the EPR of very suppressed amplitude.

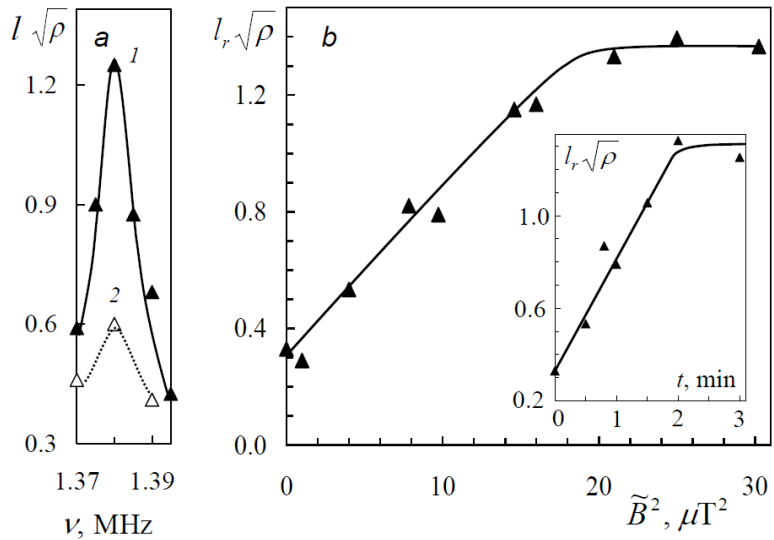

Fig. 2 (a) Peaks of the mean dislocation paths $l$ in $NaCl_{Ni}$ samples exposed for $t = 3$ min to the fields $\widetilde{\mathbf{B}} \perp \mathbf{B}_{Earth}$ (line 1) and $\widetilde{\mathbf{B}} \parallel \mathbf{B}_{Earth}$ (line 2) at $\widetilde{B} = 3.12\ \mu T$ in the vicinity of the resonance frequency 1.38 MHz of the pump field, (b) the maximum paths $l_r$ versus the square of pump field amplitude $\widetilde{B}$ (at $t = 1$ min) and the exposure time $t$ for $\widetilde{B} = 3.12\ \mu T$.

Dependence of the resonance mean path on the key parameters

Here the dependences of the resonance mean paths $l_r$ on the amplitude of the pump magnetic field $\widetilde{\mathbf{B}}$, time $t$ of exposure of samples to the magnetic field, and the concentration $C$ of the impurity is discussed. Dislocations and crossed fields are chosen mutually orthogonal in order to ensure the maximum effect.

Figure 2 (b) shows quasi-linear dependences of the mean path amplitude $l_r$ on $\widetilde{B}^2$ and $t$ with saturation at large paths, which corresponds to the depletion of relaxation of the dislocation structure. The presented results are obtained for the same $NaCl_{Ni}$ crystal as data shown in Fig. 2 (a). The linear part of these dependences can be represented by the empirical formula $l_r = l_0 + k \tilde{B}^2 t$. Here, $l_0$ is the background path caused by the etching procedure of near-surface stoppers and manipulations with the samples. The linear dependence of the increment of paths $\Delta l = l_r - l_0$ on the time $t$ means that the process of displacement of dislocations is quasi-stationary, and the proportionality $\Delta l \propto \tilde{B}^2$ likely corresponds to the quadratic dependence of the effect on the amplitude of the pump field as it is typical for the EPR phenomenon.

## 3.2 Anisotropy of the resonance frequency

The value of the resonance frequency $v_r = 1.38$ MHz of the pump field in Fig. 2 (a) is obtained from Eq. (1) for $B = B_{Earth}$ and $g = 1.97$. It was found that it does not change under the rotation of the pump field. However, the resonance frequency appeared strongly dependent on the orientation of the sample in the Earth's field. In particular, at the rotation of the sample about its edge **a** by the angle θ from the $B_{Earth}$ direction, the peak of dislocation paths shown in Fig. 2 (a) is shifted toward lower frequencies proportionally to $\cos\theta$ (see Fig. 3):

$$v_r \approx v_0 \cos\theta, \quad v_0 = g\,\mu_B B_{Earth}/h, \quad (2)$$

where $v_0 = 1.38$ MHz. Each peak specified by Eq. (2) at a fixed value is accompanied by an additional peak at the resonance frequency:

$$v'_r \approx v_0 \cos(90° - \theta) = v_0 \sin\theta. \quad (3)$$

Such a pair of peaks in Fig. 3 (a) corresponds to the angles of 30° and 60°, and these peaks at 45° are joined into one.

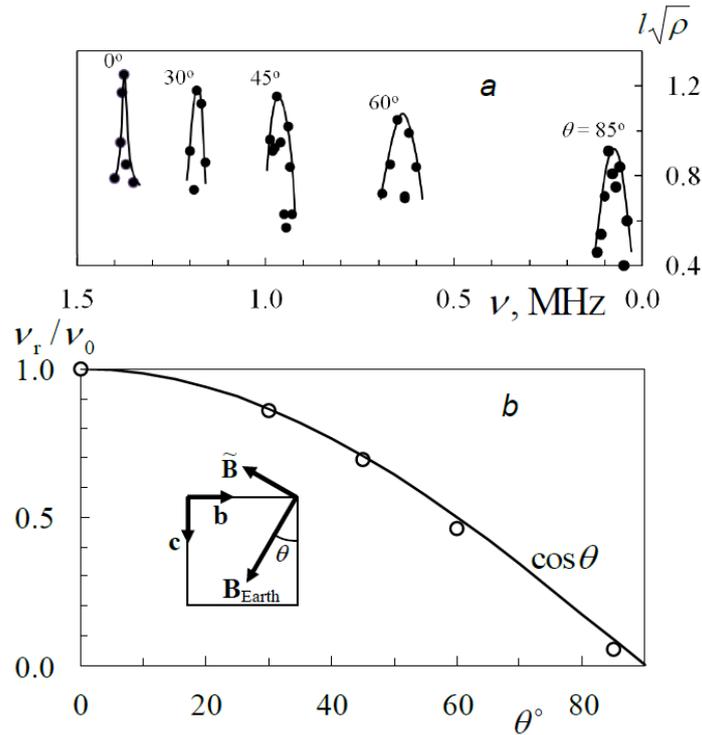

Fig. 3 (a) Peaks of resonance mean paths of **a**-dislocations for various angles θ, (b) angular dependence on the relative resonance frequency $v_r/v_0$ ($v_0 = 1.38$). The inset shows the scheme of the mutual orientation of the sample and magnetic fields.

Thus, by Eq. (2), our low-frequency EPR resonance is characterized by the giant anisotropy of the effective $g$-factor: $g_{eff} = g \cos\theta$. Interpretation of this property will be given in Section 4.

## 3.3 The resonance with a pulse pump field

There is else the other variant of the dislocation EPR, when a magnetic harmonic pumping field is replaced by a pulsed one. In this case, the resonance frequency $v_r$ in Eqn. (1) is replaced by the resonance pulse duration $\tau_r$: $v_r \to \tau^{-1}$. This effect of the influence of the Earth magnetic field was studied in a collaboration with the group of V.A. Morozov at the St. Petersburg State University Alshits (2013b). The Earth magnetic field at the place of the sample in the setup was obviously different from that in the Moscow experiments. This field was specially measured and was 66 $\mu$T at an angle to the vertical of 7.7°. Certainly, this field contained some admixture of the lab background field. The pulsed magnetic field was produced in a solenoid loaded by a rectangular pulse with the controlled duration and amplitude from the generator. Single pulses with the amplitudes $\tilde{B}_m = 4.0 \div 17.6$ $\mu$T and durations $\tau = 0.50 \div 0.57$ $\mu$T were used in the experiments described below.

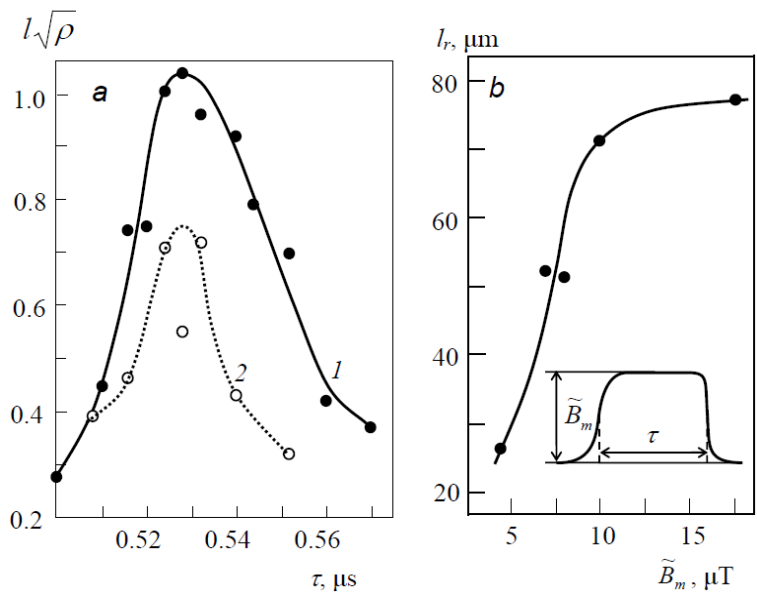

Fig. 4 (a) Normalized mean dislocation path versus the duration of the pulse, (b) the resonance maximum path versus the amplitude of the pulse for *NaCl*$_{Ca}$ crystals.

In the first series of experiments (see Fig. 4 (a)), the dependence of the mean paths $l$ of dislocations on the duration $\tau$ of the pulses of the pump field with a constant amplitude of $\tilde{B}_m = 17.6$ $\mu$T was studied. The results were obtained for the three orientations of this field, $\tilde{\mathbf{B}} \perp \mathbf{B}_{Earth}$ (1), $\tilde{\mathbf{B}} \parallel \mathbf{B}_{Earth}$ (2) and at the orientation of the sample like It was shown in Fig. 1 (b). As in the previous cases, **a**-dislocations were the most mobile. Their paths at both orientations of the pulsed pump field form pronounced resonance peaks, reaching a maximum at the same duration $\tau_r \approx 0.53$ $\mu$s, which corresponds to the *g*-factor of EPR

$$g = h/(\tau_r \mu_B B_{Earth}) \approx 2. \qquad (4)$$

In the parallel fields $\tilde{\mathbf{B}} \parallel \mathbf{B}_{Earth}$, the amplitude of the effect from which background paths are subtracted is almost half of that in orthogonal fields, which approximately corresponds to the relation for the same crystals under harmonic pumping.

At the same time, the relative path $l_r \sqrt{\rho}$ observed under the optimal conditions of the orthogonal fields is larger than unity (the absolute value is $l_r \approx 80$ $\mu$m). This indicates a high degree of relaxation of the dislocation structure, particularly taking into account that the relative density of mobile dislocations approaches 100%. This result seems striking, because it is reached in a half of a microsecond. We recall that the same degree of relaxation under harmonic pumping was reached in ~5 min.

It was assumed in Alshits (2013b) that the observed "explosive" relaxation is a collective process of self-organization of dislocations at their very fast almost simultaneous magnetically induced depinning when the ensemble becomes unstable. Such processes are usually characterized by a certain threshold of the effect. In this case, it is a threshold in the amplitude of the pulsed field. Indeed, such a threshold of the effect was experimentally revealed. As is seen in Fig. 4 (b), the resonance path $l_r$ at a decrease in the pulse amplitude begins to decrease abruptly below 10 $\mu$T, reaching background values at ~5 $\mu$T.

## 4. RESONANCE MODIFICATIONS OF MICROHARDNESS

Resonance memory effects Alshits (2012) and Alshits (2018) were studied in *ZnO* crystals grown by the hydrothermal synthesis method, as well as in triglycine sulphate and potassium acid phthalate crystals grown from aqueous solutions by the temperature reduction method.

Microhardness was measured by the Vickers method on a Neophot-21 optical microscope. Each value of microhardness was determined from the averaged size of diagonals of 20 ÷ 25 dents of the indenter. The error of these measurements was 1.5 ÷ 3%. Measurements were performed before magnetic exposure, immediately after it, each hour during the first 4 ÷ 6 h, and then 1 ÷ 3 times over the next several days.

Preliminary exposure of the crystals in crossed magnetic fields for an appropriate resonance frequency of the pump field resulted in an increase in the microhardness in the *ZnO* crystal and in its decrease in TGS and KAP crystals. The maximum change in the microhardness (10 ÷ 15%) was reached 1 ÷ 3 h after the magnetic treatment, the microhardness gradually returned to its initial value on the first day. Next, after a sufficient pause, the effect was completely reproduced under the same conditions. The delay of the response of the crystal to magnetic exposure occurs apparently because of diffusion processes after the spin transformation of impurity centres.

Figure 5 (a) shows the dependences of the relative change in the microhardness on the frequency of the pump field for all the crystals measured in definite optimal delay times after the exposure. It is seen that these dependences are clearly resonant. Magnetic memory has strong anisotropy. The effect was completely or partially suppressed when a certain direction of a crystal coincided with the Earth's magnetic field $\mathbf{B}_{\text{Earth}}$, and such a direction was found for each crystal. In the *ZnO* and TGS crystals, these are axes of 6- and 2-symmetries, respectively. This direction in the KAP crystal is the <100> direction, which lies in the cleavage plane and is orthogonal to axis 2.

It was found that the most important anisotropic property of this effect is again the strong dependence of the resonance frequency on the sample orientation in the Earth's field. As is seen from Fig. 5 (a), the three groups of peaks related to different angles of a sample inclination from the vector $\mathbf{B}_{\text{Earth}}$ are characterized by the different resonance frequencies. The simple checking shows that the corresponding relative frequencies $\nu_r / \nu_0$ are proportional to $\cos\theta$ (see Fig. 5 (b)).

Thus, in complete analogy with the dislocation effect in the *NaCl* crystal (see Fig. 3) a giant anisotropy of the effective $g$-factor: $g_{\text{eff}} \approx g \cos\theta$ was meet again. This means that the observed anomaly is a general property of the studied low-frequency EPR. As it was shown in Alshits (2018), the discussed effect probably arises due to the local magnetic fields $\mathbf{B}_{\text{loc}}$ in crystals which have definite orientation and rotates together with sample. Ordinarily, the field $B_{\text{loc}}$ is by 2-3 orders smaller than the static field $B$ in the standard EPR, so it produces the ordinary small anisotropy of the $g$-factor. However, in our case and the situation is radically different. According to Alshits (2018) the Zeeman's splitting of the magnetic energy level is determined by

$$h\nu_r = g\mu_B(|\mathbf{B}_{\text{Earth}} + \mathbf{B}_{\text{loc}}| - \mathbf{B}_{\text{loc}}), \qquad (5)$$

where for $B_{\text{loc}} \gg B_{\text{Earth}}$ one has

$$|\mathbf{B}_{\text{Earth}} + \mathbf{B}_{\text{loc}}| - B_{\text{loc}} \approx \sqrt{B_{\text{loc}}^2 + 2\mathbf{B}_{\text{Earth}} \cdot \mathbf{B}_{\text{loc}}} - B_{\text{loc}} \approx B_{\text{Earth}}\cos\theta. \qquad (6)$$

Thus, we come to the observed dependence descby equation (2) which may be described by the effective $g$-factor $g_{\text{eff}} \approx g \cos\theta$.

.

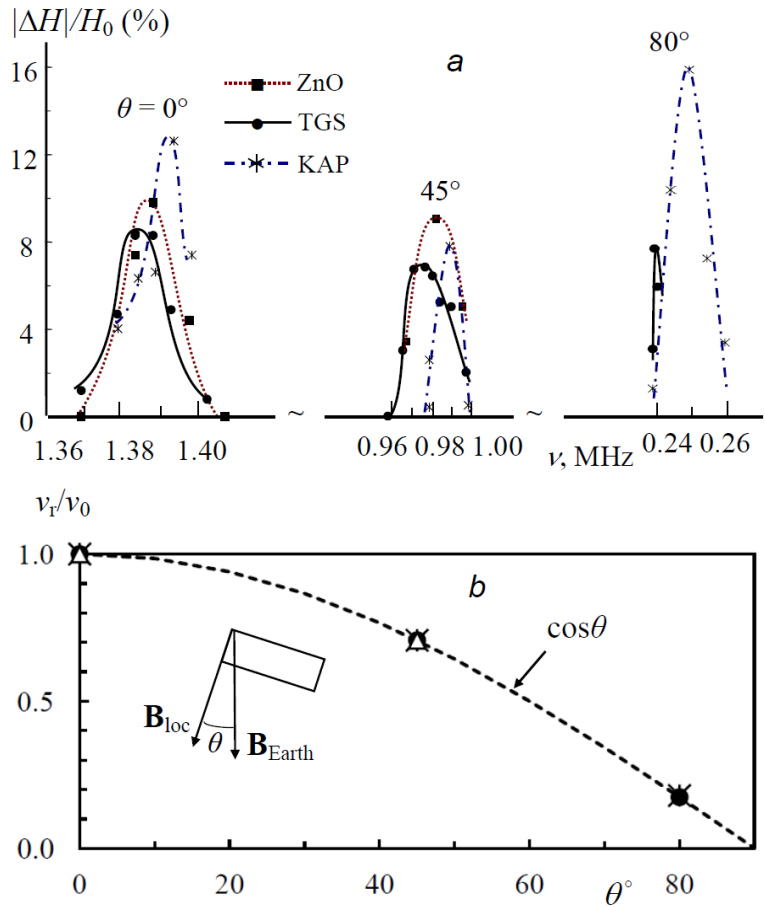

Fig. 5 (a) Resonance peaks of the microhardness changes $|\Delta H|/H_0$ for *ZnO*, TGS and KAP crystals at three angles $\theta$ of sample rotations by 0, 45° and 80° about the $\mathbf{B}_{Earth}$ direction, (b) the relative resonance frequency $\nu_r / \nu_0$ versus the angle $\theta$ of sample rotation.

## 5. CONCLUSIONS

    The features of the resonance magnetoplasticity in ultralow magnetic fields have been studied primarily for *NaCl* crystals, which have been used for a long time as a convenient model objects for the extensive studies of various properties of the magnetoplastic effect. This work concerns a new direction of studies where the exposure of crystals occurs in the EPR scheme with the use of the Earth's magnetic field and radiofrequency pumping.

    It has been found that the observed resonance of dislocation mobility has new, very specific, strongly anisotropic properties. The main such feature is the sensitivity of the mean path and the resonance frequency to the orientation of the sample and the directions of dislocations with respect to crossed magnetic fields. This sensitivity was not previously observed under usual EPR conditions corresponding to the pump frequencies of 10 GHz and 150 MHz.

    Pulse EPR has also been detected in *NaCl* crystals in the Earth's magnetic field. The resonance peak of paths appears in a narrow range of durations of a pump field pulse near 0.53 $\mu s$ (see Fig. 4 (a)). At fairly large amplitudes of the pump pulse beginning with a threshold level of ~10 $\mu T$ (see Fig. 4 (b)), an explosive coherent relaxation of the dislocation structure occurs: almost all fresh dislocations move in 0.5 $\mu s$ to the same distances of ~100 $\mu m$ as under harmonic pumping in a time of exposure of ~5 min.

    The resonance change in the microhardness of the *ZnO*, TGS, and KAP crystals after their exposure in the same EPR scheme in the Earth's magnetic field indicates that this phenomenon and its specific anisotropic features are fairly general. It occurs in the very different materials and is responsible for both

in situ effects and magnetic memory effects. In essence, this concerns a very wide range of relaxation processes that have magnetic resonance nature.

An important consequence of the existence of magnetic resonance transformations of defects is the recognition of new dangers of modern technical civilization. The mentioned processes can occur uncontrollably in critical elements of constructions and devices, leading to their degradation, when weak background radiofrequency fields are imposed on the Earth's magnetic field. These are not abstract possibilities but are real processes, e.g., the accidental observation in 1985 of mass displacements of dislocations in a *NaCl* crystal under the action of magnetic pickups because of click of a switch described in Alshits (1999). This observation initiated the study of magnetoplasticity.

This work was partly supported by the Federal Agency of Scientific Organizations of Russia (Agreement No. 007-GZ/Ch3363/26) and by the Presidium of the Russian Academy of Sciences (program no. 32 for 2018-2020). Authors are grateful to V.A. Morozov and S.A. Minyukov for collaboration and helpful remarks.